\documentclass[twocolumn,aps,pra,amsmath,amssymb,showpacs]{revtex4}
\usepackage{float}
\usepackage{amsmath}
\usepackage{amssymb}
\usepackage{graphicx}
\usepackage{esint}
\usepackage{color}

\makeatletter

\@ifundefined{textcolor}{}
{%
 \definecolor{BLACK}{gray}{0}
 \definecolor{WHITE}{gray}{1}
 \definecolor{RED}{rgb}{1,0,0}
 \definecolor{GREEN}{rgb}{0,1,0}
 \definecolor{BLUE}{rgb}{0,0,1}
 \definecolor{CYAN}{cmyk}{1,0,0,0}
 \definecolor{MAGENTA}{cmyk}{0,1,0,0}
 \definecolor{YELLOW}{cmyk}{0,0,1,0}
}

\usepackage{epsfig}\usepackage{amsfonts}

\makeatother

\begin{document}

\title{Enhanced absorption microscopy with correlated photon pairs}

\author{Ming Li, $^{1,2,\dagger}$ Chang-Ling Zou, $^{1,2,\dagger}$ Di Liu,
	$^{1,2}$ Guo-Ping Guo, $^{1,2}$ Guang-Can Guo $^{1,2}$ and Xi-Feng
	Ren$^{1,2,}$}
\email{renxf@ustc.edu.cn}

\affiliation{$^{1}$Key Lab of Quantum Information, University of Science and
	Technology of China, CAS, Hefei, Anhui, 230026, China.}

\affiliation{$^{2}$Synergetic Innovation Center of Quantum Information $\&$
	Quantum Physics, University of Science and Technology of China, Hefei,
	Anhui 230026, China}

\begin{abstract}
By harnessing the quantum states of light for illumination, precise
phase and absorption estimations can be achieved with precision beyond
the standard quantum limit. Despite their significance for precision
measurements, quantum states are fragile in noisy environments which
leads to difficulties in revealing the quantum advantage. In this
work, we propose a scheme to improve optical absorption estimation
precision by using the correlated photon pairs from spontaneous parametric
down-conversion as the illumination. This scheme performs better than clasical illumination when
the loss is below a critical value. Experimentally, the scheme
is demonstrated by a scanning transmission type microscope that uses
correlated photon illumination. As a result, the signal-to-noise ratio
(SNR) of a two-photon image shows a $1.36$-fold enhancement over
that of single-photon image for a single-layer graphene with a $0.98$
transmittance. This enhancement factor will be larger when  using multi-mode squeezed state as the illumination. 
\end{abstract}
\maketitle

\section{Introduction}

Low-energy illumination imaging is necessary
when treating photosensitive samples, including biological cells \citep{biology sample,phyrep2016}
, quantum gases \citep{quantumgas}, and atomic ensembles \citep{atomensenble}.
The qualities of the images can be measured by their signal-to-noise
ratios (SNRs), where the signal is the contrast between the sample
and background and the noise is the fluctuation of the detected signal.
According to the theory of statistics \citep{book1}, the estimation
precision $\Delta t$ is limited by $\Delta t\geqslant\frac{1}{\sqrt{MF(t)}}$,
where $M$ is the number of repeated measurements, $F(t)$ is the
value of the Fisher information. For low-photon-level coherent state
illumination, the shot-noise limit appears to be dominant, and the
measurement precision can be improved by repeating the measurements
\citep{SQL}. Fortunately, quantum metrology \citep{metrologyscience,squeeze1,squeeze2,hybrid,metrology,Simon2017}
further improves the parameter estimation precision \citep{pryde 2017}.
By carefully choosing the probe quantum state and projection detection
basis, the Fisher information of single measurement could be improved,
and the Heisenberg limit of the $\sqrt{N}$ enhancement of the Fisher
information can be potentially approached \citep{HL}. It has already
been demonstrated that the quantum $N00N$ state \citep{noon silberger}
can be used to measure the optical phase such that the measurements
precision exceeds the standard quantum limit \citep{metrology2,prl1990,dbwavelength,dbprl,science2007,njpnp,xiang,xiang 2}.
Recently, such methods were introduced to build microscopes with quantum
light illumination that outperform the same type of interference microscope
with classical illumination \citep{nc2013,prl2014}. However, such
methods have limited applications and depend on the target samples,
additionally, the quantum entanglement is fragile in a noisy environment.
The apparatus also requires a high stability for interference. 

For direct absorption measurements, theoretical and experimental works
have demonstrated that the heralded single-photon source from parametric
down-conversion \citep{xiaosongma} can exceed the shot-noise limit
when estimating the transmittance of highly transparent samples \citep{photonnumber}.
In the protocol named ``quantum illumination'', the heralded single
photons are sent to a low reflectivity target in a high background-noise
environment, and an enhanced SNR for the detection can be obtained
\citep{QI,QI2,QI3,QI4,QI5,QI6}. Recently, Matthews and coworkers
experimentally demonstrated that the sub-Poisson distribution of quantum
states can reduce the noise when estimating lossy samples \citep{njp2017,arxivabsolute,prappl}.
Their work used a high-quality heralded single-photon source from
a spontaneous down-conversion (SPDC) process to measure the transmittance
beyond the classical shot-noise limit. Such a method is suitable for
the commonly used transmission or reflection microscopes and does
not rely on quantum entanglement or quantum interference. However, the quantum advantages are sensitive to 
the heralding efficiency of the single-photon source, which is limited by
the optical loss and inefficient detector in the reference arm. Unfortunately, the quality of the
heralded probe single-photon source is far from perfect in most situations.

In this letter, we propose a scheme to obtain an enhanced precision
absorption microscope by sending both correlated photons to the sample to probe the transmittance. Our scheme can surpass the coherent state limit without heralding apparatus when the transmittance is near unity. Above a criticle value of the transmittance, it performs better than the classical illumination in the same measurement system.
The enhancements are demonstrated experimentally with a scanning microscope, achieving
an enhancement factor of $1.36$ for a graphene sample transparency
of $0.98$. Using direct measurements, the scheme is stable and could
be used for applications in photosensitive biological imaging. 

\section{Principle}

For a directly absorption measurement, the probability
of the input photon passing the sample is a function of the transmittance
$t$. Representing the bosonic operator for input and output photon
modes as $a_{i}$ and $b_{i}$, respectively, the linear absorption
can be effectively treated as a beam splitter (BS) with transmission
$t$, and thus, $b_{i}=\sqrt{t}a_{i}+\sqrt{1-t}c$, where $c$ is
the operator of the vacuum noise input mode. By measuring the expectation
value of an appropriate physical observable $O$ as a function of
$b_{i}$, we can estimate the transmittance $t$ of the absorptive
sample such that the uncertainty is written as
\begin{equation}
\Delta t_{M}=\frac{\sqrt{\langle\Delta O^{2}\rangle}}{\sqrt{M}|\frac{\partial\langle O\rangle}{\partial t}|},
\end{equation}
where the fluctuation $\langle\Delta O^{2}\rangle=\langle O^{2}\rangle-\langle O\rangle^{2}$
and $M$ is the number of repeated measurements. Generally, $O$ is
chosen to be an expression of the photon number operators that correspond
to the photon coincidence detection, which can be easily realized
experimentally. 

Fig.$\,$1(a) shows two schemes for the absorption
measurement. A sequence of correlated photon pairs is generated and
then sent to the observers, Alice and Bob. In the left case, only
the signal photons sent to Bob pass through the absorptive sample;
this case is called the ``\emph{single-pass}'' scheme. In the right
case, both the signal and idle photons pass through the sample; this
case is called the ``\emph{double-pass}'' scheme. Similar to quantum
entanglement-assisted dense coding, we question whether the correlation
can obtain more information in the ``\emph{double-pass}'' scheme
than that in the ``\emph{single-pass}'' scheme, with the same illumination
photon number. Therefore, we compare the two schemes shown in Fig.$\,$1(a),
where the photon source is a two-mode squeezing state with $\left|\phi\right\rangle =\alpha\sum_{n=0}^{\infty}\beta^{n}\left|n,n\right\rangle $,
corresponding to the non-degenerate down-conversion sources, such
that $\beta$ is related to the squeezing parameter and $\alpha$ is the normalization factor. For both cases, we perform
the $k_{1}$-th and $k_{2}$-th order correlation measurements of
each mode, i.e., the physical operator we measure can be expressed
as $O=(b_{1}^{\dagger})^{k_{1}}b_{1}^{k_{1}}(b_{2}^{\dagger})^{k_{2}}b_{2}^{k_{2}}$
with the subscript $1(2)$ denotes the path to Alice (Bob). Let the
transmittances be $t_{1}$ and $t_{2}$ , we then have $\langle O\rangle=t_{1}^{k_{1}}t_{2}^{k_{2}}\langle(a_{1}^{\dagger})^{k_{1}}a_{1}^{k_{1}}(a_{2}^{\dagger})^{k_{2}}a_{2}^{k_{2}}\rangle$
and $\langle O^{2}\rangle=\sum_{m,l}^{k_{1},k_{2}}C{}_{k_{1},m}C_{k_{2},l}t_{1}^{k_{1}+m}t_{2}^{k_{2}+l}\langle(a_{1}^{\dagger})^{k_{1}+m}a_{1}^{k_{1}+m}(a_{2}^{\dagger})^{k_{2}+l}a_{2}^{k_{2}+l}\rangle$,
where $C_{k,m}$ is the parameter transforming $O^{2}$ to normal
order operators. Introducing
the precision of the absorption estimation for $t_{2}$ 
\begin{equation}
\Delta t=\sqrt{R}\frac{\sqrt{\langle\Delta O^{2}\rangle}}{|\frac{\partial\langle O\rangle}{\partial t}|},
\end{equation}
where $R=\sum_{j}\langle\phi|b_{j}^{\dagger}b_{j}|\phi\rangle$ is
the averaged photon number of the input state ($j=1$ for ``\emph{single-pass}''
scheme, $j=1,2$ for ``\emph{double-pass}'' scheme). Therefore,
$\Delta t$ can also be understood as the normalized precision of
the estimation per input photon.

\begin{figure}
	\includegraphics[width=8cm]{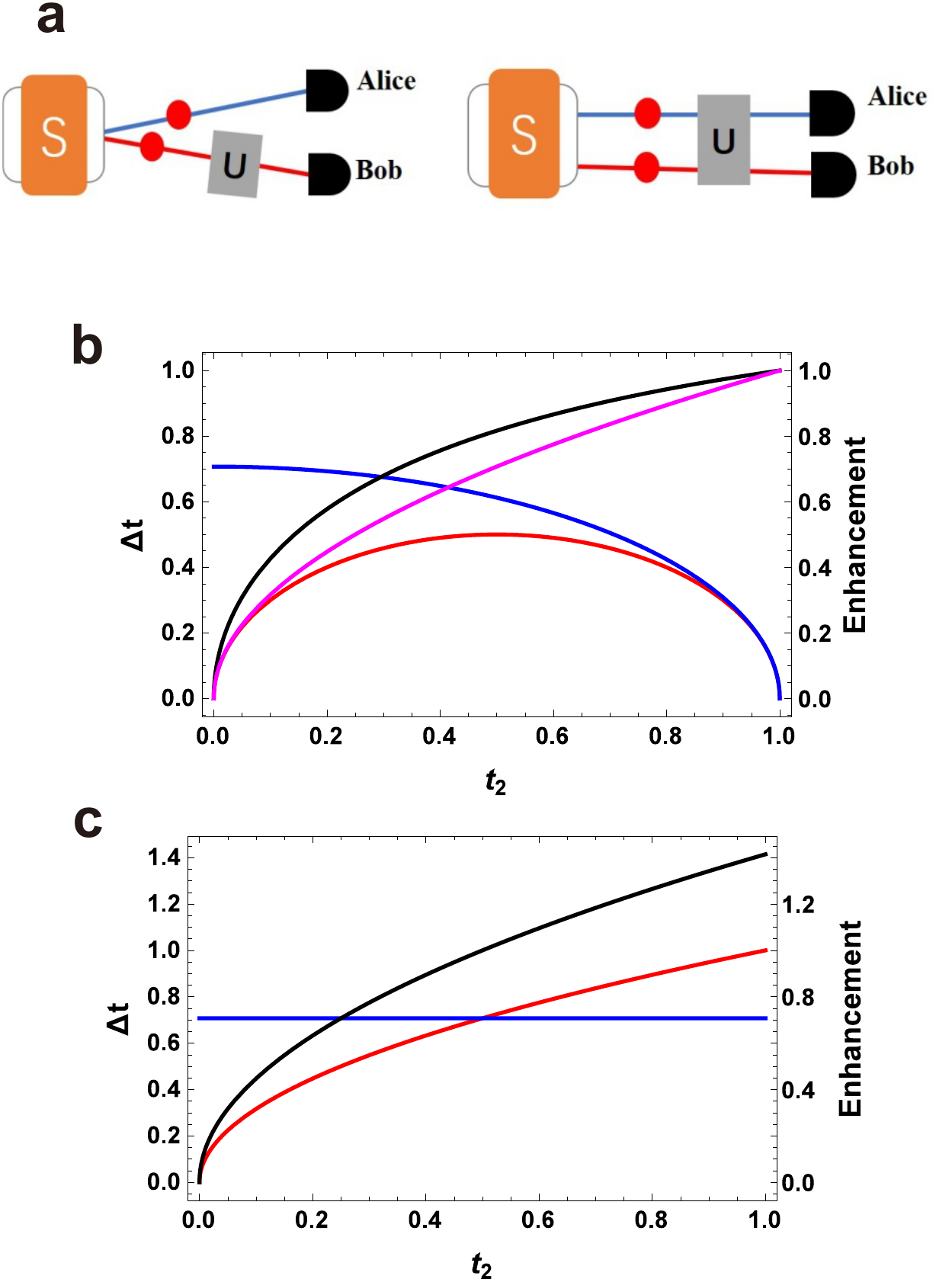}
	
	\caption{Theoretical calculations of the transmittance estimates for different
		illumination schemes. (a) Schematic pictures of the ``\emph{single-pass}''
		(left) and the ``\emph{double-pass}'' schemes (right). (b) Transmittance
		estimate precision for an ideal single-photon probe state. (c)Transmittance
		estimate precision for the probe state from SPDC. In this case, the
		input state has fluctuations of the photon number. Here, we set $\beta=10^{-2}$.
		The estimation precision of the two-photon case exceeds that of the
		single-photon case when $t_{2}$ is greater than the critical value
		$t_{0}$. Red line: ``\emph{single-pass}'' scheme. Blue line: ``\emph{double-pass}'' scheme. Pink line: coherent state. Black line: enhancement
		factor $\Delta t_{\mathrm{SP}}/\Delta t_{\mathrm{DP}}$.}
	
	\label{Principle}
\end{figure}

For the simplest case, we choose an ideal photon pair state $\left|\phi\right\rangle =\left|1,1\right\rangle $
and $k_{1}=k_{2}=1$. The precision $\Delta t_{\mathrm{SP}}$ for
the ``\emph{single-pass}'' scheme ($t_{1}=1$), $\Delta t_{\mathrm{DP}}$
for the ``\emph{double-pass}'' scheme and $\Delta t_{\mathrm{CS}}$
for a coherent state input are calculated as follows:
\begin{eqnarray}
\Delta t_{\mathrm{SP}} & = & \sqrt{t_{2}(1-t_{2})},\\
\Delta t_{\mathrm{DP}} & = & \sqrt{(1-t_{2}^{2})/2},\\
\Delta_{\mathrm{CS}} & = & \sqrt{t_{2}}.
\end{eqnarray}
These precision as well as the enhancement factor $\Delta t_{\mathrm{SP}}/\Delta t_{\mathrm{DP}}$
are plotted in Fig.$\,$\ref{Principle}(b). It is shown that both
schemes outperform the classical case of the coherent state for high
transmittance and that the ``\emph{single-pass}'' scheme shows the
best performance for all transmittance values, which is the same with single photon state. For such an ideal input
quantum state, the minimal photon number uncertainty of the quantum
state results in the suppression of the noise in the direct absorption
parameter estimation. In this case, the ``\emph{double-pass}'' scheme
shows no advantage over the ``\emph{single-pass}'' scheme.

However, the advantage of single photons requires high-performance heralding apparatus on the SPDC source and the imperfections in the operation and detection will degrade the quantum advantage. The ``\emph{double-pass}'' scheme can reveal the quantum advantage without making heralding on the SPDC source. Usually, the SPDC source works with
$\beta\ll1$ to avoid multiple photon pair generation, generating
a state that can be approximated as $|\phi\rangle=\alpha|0,0\rangle+\beta|1,1\rangle$.
For such a practical source, 
\begin{align}
\Delta t_{\mathrm{SP}} & =\sqrt{t_{2}-t_{2}^{2}\beta^{2}}\approx\sqrt{t_{2}},\label{eq:sp}\\
\Delta t_{\mathrm{DP}} & =\sqrt{(1-t_{2}^{2}\beta^{2})/2}\approx1/\sqrt{2}.\label{eq:dp}
\end{align}
The equations indicate that the estimation precision of ``\emph{single-pass}'' scheme
approximately equals $\Delta t_{CS}$ and the estimation precision
of ``\emph{double-pass}'' scheme is almost independent on the transmittance.
So, the precision of ``\emph{single-pass}'' can be approximately
treated as the classical bound for coherent state illumination in the same apparatus. In Fig.$\,$\ref{Principle}(c),
we use the estimated value from experiment $\beta=10^{-2}$ and calculate
the precisions $\Delta t_{\mathrm{SP}}$ and $\Delta t_{\mathrm{DP}}$
as well as the enhancement factor $\Delta t_{\mathrm{SP}}/\Delta t_{\mathrm{DP}}$.
It is shown that the ``\emph{double-pass}'' scheme outperforms the
``\emph{single-pass}'' scheme when $t_{2}$ is higher than the critical
value $t_{\mathrm{critical}}$. The maximum enhancement factor is
$\sqrt{2}$ when $t_{2}$ approaches unity. The enhancement can be understood by Eq. (2). For both schemes, the probe state and the detection are the same. The differences of the ``\emph{single-pass}'' and ``\emph{double-pass}'' schemes are two folds: (i) The main difference between the two schemes is that they measure different quantities. The ``\emph{single-pass}'' scheme measures $t$ while the ``\emph{single-pass}'' scheme measures $t^{2}$. The sensitivity increases by a factor 2 for the later scheme. (ii) When the transmittance of the sample approaches to unity, the means and variances of the coincidence detection rates in the two schemes are nearly the same. The average number of photons passing through the sample in the ``\emph{double-pass}'' scheme is twice of that in the ``\emph{single-pass}'' scheme, resulting the decrease of the sensitivity by a factor of $\sqrt{2}$. Therefore, the precision is finally enhanced by $\sqrt{2}$ times when the sample is nearly transparent. By solving the equation
$\Delta t_{\mathrm{SP}}=\Delta t_{\mathrm{DP}}$, we derive the relation
between the critical point $t_{\mathrm{critical}}$ and $\beta$ as
\begin{equation}
t_{\mathrm{critical}}=\frac{1-\sqrt{1-\beta^{2}}}{\beta^{2}}\approx\frac{1}{2}+\frac{1}{8}\beta^{2}.
\end{equation}
Our analysis indicates that the ``\emph{double-pass}'' scheme can be
used to suppress the probe state fluctuation and improve the precision
of the estimation without performing additional heralding on the photon
source, when the sample transmittance exceeds a critical value $t_{\mathrm{critical}}$.
Throughout our analysis, we have assumed the detection efficiency
to be unity. For a non-ideal detection process, the inefficiency of
the detectors can be summarized as the loss of the sample and does not affect the validity of the calculations.

\section{Experimental results}

\begin{figure}
	\includegraphics[width=8cm]{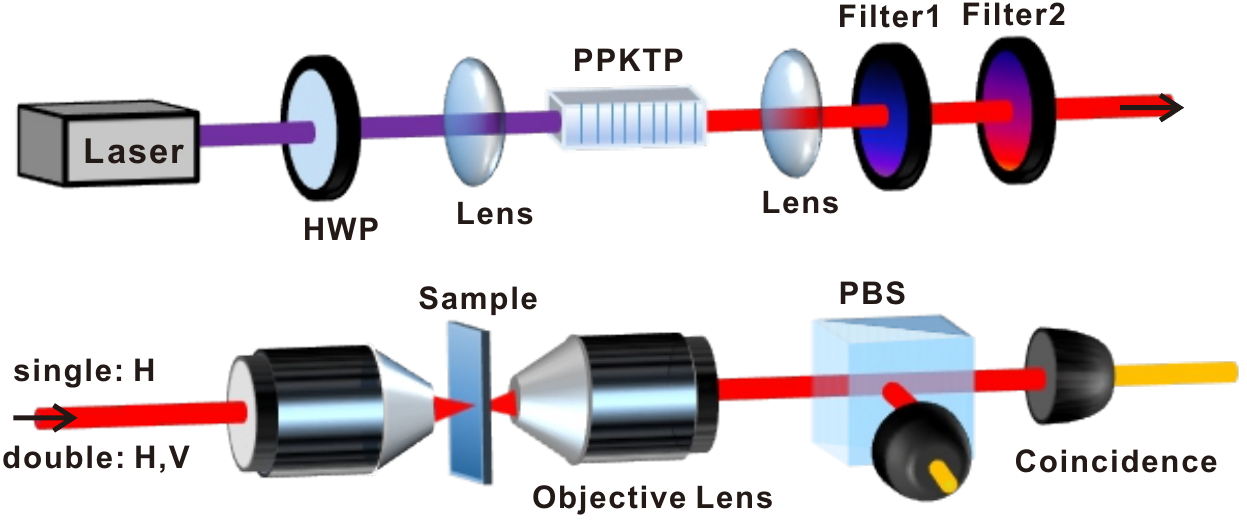}
	
	\caption{Experimental setup. HWP: half-wave plate; Filter1: $650\,\mathrm{nm}$
		long-pass filter; Filter2: $808\,\mathrm{nm}$ interference filter.
		PBS: polarization beam splitter. A $404\,\mathrm{nm}$ laser is focused
		on the periodic pooled potassium titanyl phosphate (KTP) crystal to
		generate twin photons with orthogonal polarizations ($H,V$). We use
		a temperature controller to tune the wavelength of the photons. ``\emph{Single-pass}''
		scheme: After filtering and PBS, the photons are separated and coupled
		into different single mode fibers. One photon is used as the trigger,
		and the other is used as the illumination. ``\emph{Double-pass}''
		scheme: After filtering, the two photons are coupled to one fiber
		and then the two-photon source is used as the illumination and is
		focused on the sample. The coincidence measurement is performed by
		splitting the photons with the PBS. The difference between the two
		schemes is the position of the PBS. The ``\emph{Single-pass}'' scheme
		is realized by moving the PBS in the black dashed box upwards.}
	
	\label{Setup}
\end{figure}

In this section, we experimentally carry
out both ``\emph{single-pass}'' and ``\emph{double-pass}'' absorption
microscope measurements and give a proof of principle demonstration
of the advantage of the ``\emph{double-pass}''
scheme. The experimental setup is shown in Fig.$\,$\ref{Setup}. A
periodic pooled KTP crystal is pumped by a $404\,\mathrm{nm}$ continuous-wave
laser, and the wavelength degenerate photon pairs with orthogonal
polarizations at $808\,\mathrm{nm}$ are generated via a type-II SPDC
process. To satisfy the quasi phase-matching condition \citep{qpm Sergienko}
for the wavelength degenerate SPDC, the nonlinear crystal is put in
a temperature controlled oven. Filtered by a $650\,\mathrm{nm}$ long-pass
filter and $808\,\mathrm{nm}$ interference filter, photon pairs are
separated from the pump and are further divided by a polarization
beam splitter (PBS) before been finally coupled into different single-mode
fibers before the sample (for ``\emph{single-pass}'' scheme) or
after the sample (for ``\emph{double-pass}'' scheme). 

The scanning absorption microscope is composed of a fiber-collection
lens and two objective lenses, with N.A.s of $0.75$ and $0.8$, respectively.
The test samples to are few-layer graphene films. The illumination
photons are focused on the sample by the first objective lens, and
the transmitted photons are collected by the second objective lens
behind the sample. There are two reasons for choosing a few-layer
graphene film as the sample: (a) The transmittance is nearly unity
for a monolayer; (2) The thickness of the sample is quantized and
uniform in different areas, which gives a distribution of the transmittance
for the scanning measurement.

\begin{figure}
	\includegraphics[width=8cm]{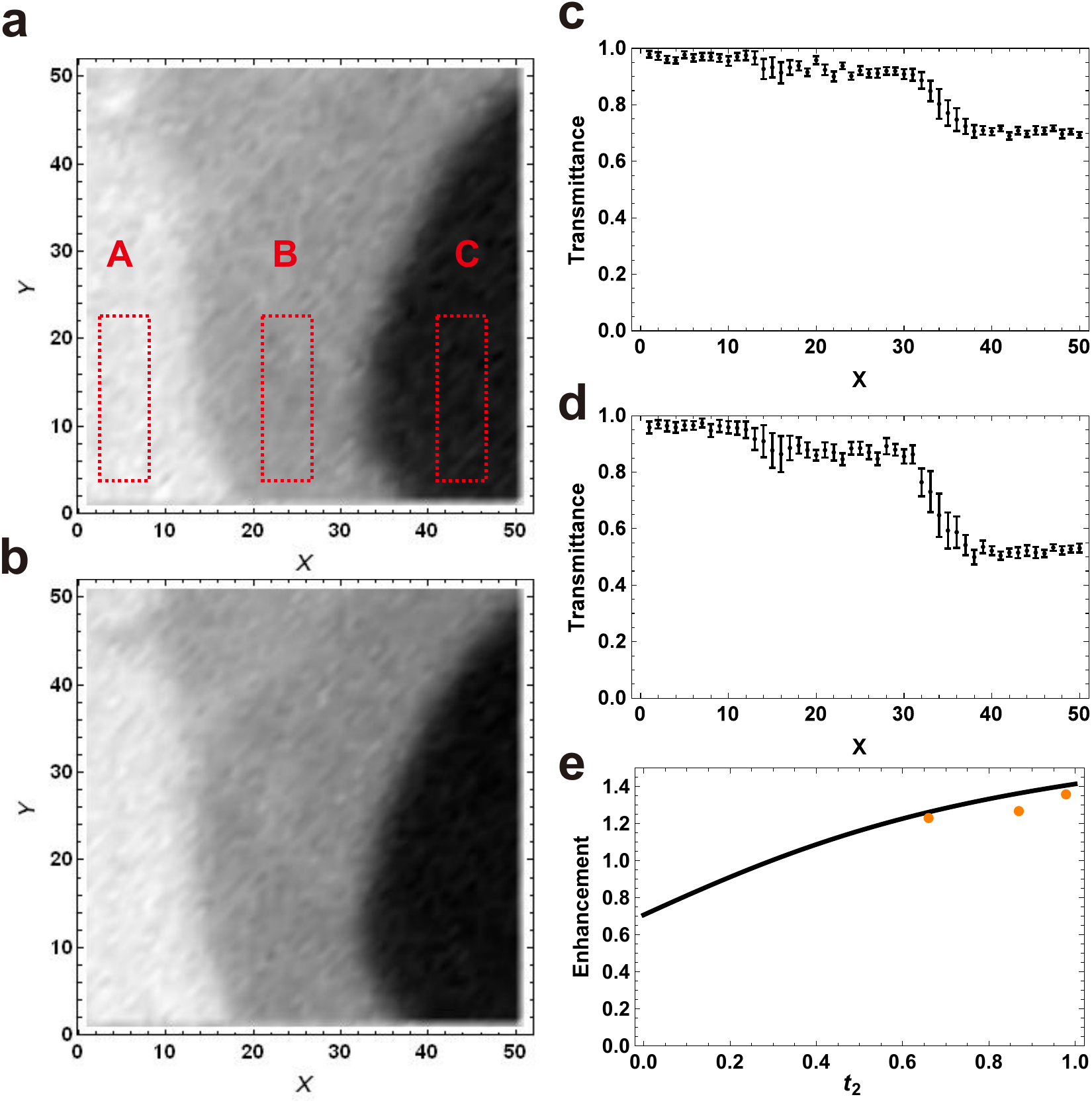}
	
	\caption{Scanned images of graphene layers from the transmission microscope.
		(a) \textquotedblleft \emph{Single-pass}\textquotedblright{} scheme.
		(b) ``\emph{Double-pass}'' scheme. Each pixel in these images corresponds
		to $100\,\mathrm{nm}$. The whole area is $5\,\mathrm{\mu m}\times5\,\mathrm{\mu m}$.
		The data used for calculating the enhancement are the counts in the
		three different areas. In the experiment, the photon fluxes are fixed
		to approximately $5000$ counts. The $SNR$s and enhancements of the
		samples with different transmittances are calculated from the data
		in the red dashed rectangles. (c)-(d) The data are extracted from
		the images shown in Fig.3. The error bars are calculated from $25$
		datasets. (e) The relation between the maximal enhancement of the
		$SNR$ and the estimated transmittance $t_{2}$. Black line: Enhancement
		of the ``\emph{double-pass}'' scheme compared to that of ``\emph{single-pass}''
		scheme. Orange points: experimental results for the graphene layers
		with different transmittances. From left to right, the transmittances
		are $0.66$, $0.87$ and $0.98$. }
	
	\label{2Dimage}
\end{figure}

Both schemes shown in Fig.$\,$\ref{Principle}(a) are performed in
our experiment. The scanned images are shown in Figs.$\,$\ref{2Dimage}(a)
and (b), in which each pixel is measured by recording the coincidence
counts within a certain period. It should be noted that the durations
of the experiments for two schemes are different to ensure the same
photon numbers. There are three distinct areas of different transmittances
in the sample that correspond to different numbers of layers. In our
experiment, we make a post-selection by adjusting the wait times in
each scheme in area $A$ to remove the influences of other losses
within the apparatus and to ensure that the detected photon numbers
are the same. Therefore, the illumination state after the area $A$
can be treated as the input. To characterize the quality of the image,
we calculate the SNRs in area $B$ and area $C$, both compared with
area $A$, whose averaged photon number is approximately $5000$ counts.
In the following experiments, the signal is defined as the average
photon number difference between the input and output counts $\langle O_{in}\rangle-\langle O_{out}\rangle$,
where $O_{in}=a_{1}^{\dagger}a_{1}a_{2}^{\dagger}a_{2}$, $O_{out}=b_{1}^{\dagger}b_{1}b_{2}^{\dagger}b_{2}$.
The uncertainty of the counts is $\langle\Delta O^{2}\rangle=\sqrt{\langle\Delta O_{in}^{2}\rangle+\langle\Delta O_{out}^{2}}\rangle$,
where $\langle\Delta O_{in}^{2}\rangle$ and $\langle\Delta O_{out}^{2}\rangle$
can be calculated from the standard deviations of the counts in each
pixel. The $SNR$ is defined as 
\begin{equation}
SNR=\frac{\langle O_{in}\rangle-\langle O_{out}\rangle}{\sqrt{\langle\Delta^{2}O_{in}\rangle+\langle\Delta^{2}O_{out}}\rangle},\label{eq:snr}
\end{equation}
which is the ratio of the expectation value of $O_{in}-O_{out}$ and
the square root of its variance.

For the \textquotedblleft \emph{single-pass}\textquotedblright{} scheme,
the idler photon is used as a trigger since it can eliminate the influence
of the environment noise, and the signal photon is used to illuminate
the sample. The measured transmittance of area $B$ and $C$ are $0.87\pm0.015$
and $0.66\pm0.013$, respectively. The calculated SNRs for $B$ to
$A$ and $C$ to $A$ are $9.99$ and $25.96$, respectively. Then,
we studied the ``\emph{double-pass}'' scheme by changing the illumination source
from a heralded single-photon to a correlated photon-pair source.
This change was realized by directly coupling the photon pairs into
a polarization-calibrated single-mode fiber. Therefore, both photons
are illuminated on the sample, and the coincidence detection is realized
with the assistance of a polarization beam splitter. Since there is
no classical or quantum interference between the photons in our scheme,
the microscope is robust to the optical path length fluctuations.
We then perform the same scanning process for the ``\emph{single-pass}''
case. To ensure the used resources are the same for the two schemes,
we adjust the measurement time to match that of the ``\emph{single-pass}''
case. The average coincidence count is approximately $2480$ counts,
approximately equal to that of the single-photon count. The measured
transmittance of area $B$ and $C$ are in ``\emph{double-pass}''
scheme are $0.87\pm0.012$ and $0.66\pm0.011$, respectively. Fig.$\,$\ref{2Dimage}(b)
shows the scanned image, which shows a higher contrast than the single-photon
case. Fig.$\,$\ref{2Dimage}(c) and (d) show detailed experimental
results, including the uncertainties of the measured transmittances
in different areas. In addition to the intensity image, we make a
normalization of the data from the $25$ lines of pixels from the
downside. The qualities of the images can now be approximately described
by the ratio between the height of the ladder and the error bar. From
our experimental data and theoretical analyses, we find that the enhancement
is relevant to the relative transmittances of the sample. The calculated
SNRs of the images for $B$ to $A$ and $C$ to $A$ are $13.18$
and $30.54$, respectively, which are higher than those of the ``\emph{single-pass}''
case. The enhancement factors are $1.268$ and $1.231$, respectively.
We have also measured a single-layer graphene film with a transmittance
of $0.98$, resulting a $1.36$-fold enhancement of the $SNR$. These
experiment results ambiguously demonstrate the enhancement of the
absorption microscope with the use of ``\emph{double-passed}'' correlated
photons. 

We plotted the $SNR$ enhancements derived from Eq.$\,$\ref{eq:snr}
for different schemes in Fig.$\,$\ref{2Dimage}(e). From the plot,
the higher transmittance gives a greater enhancement, with an upper
bound of $\sqrt{2}$ when the sample is nearly transparent. In our
experiment, the enhancement factors $1.36$, $1.27$ and $1.23$ correspond
to the transmittance of $0.98$, $0.87$ and $0.66$, respectively,
which well fit our analyses.

\section{discussion}

For realistic applications, it is important to
optimize the efficiencies of every component of the microscope, including
the input quantum photon source and the detector efficiencies. Any
losses in the state preparation, evolution and detection processes
can degrade the photon correlation, and the advantage of the ``\emph{double-pass}'' scheme is only valid when the whole transmittance is greater
than the critical transmittance $t_{\mathrm{critical}}$. An possible
method of achieving greater enhancements is utilizing the higher-order
correlations of multiphoton input probe states. A theoretical analysis
shows that the maximum enhancement over the classical bound (``\emph{single-pass}''
scheme) can be promoted to $\sqrt{N}$ by using $N$-photon correlated illumination state. 

\section{Conclusion}

In summary, we propose and experimentally demonstrate
the ``\emph{double-pass}'' scheme for direct absorption measurements
using correlated two-photon source, which achieves a higher precision
and outperforms the ``\emph{single-pass}'' case with the same resource
consumption. This method can be used to achieve low noise images in transmission or reflection microscopy better than the coherent state illumination without building high-quality heralded single photon sources.
Such transmission microscopes are very robust for practically applications
since they have no specific requirements for the stability of their
environments, unlike those required for the interference protocols.
Another advantage of this method is that the coincidence technology
is immune to optical background noise, which is quite important for
low photon number illumination measurements. The proof of principle
demonstration of this method opens a new avenue for weak field imaging
without fragile quantum interference and is promising for the practical
applications of quantum enhanced imaging.

\begin{acknowledgments}
This work was supported by the National Natural Science Foundation
of China (Nos. 61590932, 61505195 and 11774333), Anhui Initiative
in Quantum Information Technologies(No. AHY130300), the Strategic
Priority Research Program of the Chinese Academy of Sciences (No.
XDB24030601), the National Key R \& D Program (No. 2016YFA0301700),
and the Fundamental Research Funds for the Central Universities. This
work was partially carried out at the USTC Center for Micro and Nanoscale
Research and Fabrication.
\end{acknowledgments}

\section{Appendix}
The ``\emph{double-pass}'' scheme can be extended to the  ``\emph{multiple-pass}''
scheme with multiple correalted photons impinging on the sample. The physical
observable is chosen as 
\begin{equation}
O=\prod_{i}^{N}(b_{i}^{\dagger})^{k_{i}}b_{i}^{k_{i}},
\end{equation}
corresponding to the $N$ photon coincidence measurement. The expectation
values of $O$ and $O^{2}$ at the detector are
\begin{eqnarray}
\langle O\rangle & = & \prod_{i}^{N}t_{i}^{k_{i}}\langle\prod_{i}^{N}(a_{i}^{\dagger})^{k_{i}}a_{i}^{k_{i}}\rangle\\
\langle O^{2}\rangle & = & \sum_{m_{1}}^{k_{1}}\sum_{m_{2}}^{k_{2}}...\sum_{m_{N}}^{k_{N}}\prod_{i}^{N}C_{k_{i},m_{i}}t_{i}^{k_{1}+m_{i}}\langle\prod_{i}^{N}(a_{i}^{\dagger})^{k_{i}+m_{i}}a_{i}^{k_{i}+m_{i}}\rangle\nonumber \\
\end{eqnarray}
where $k_{i}$ is the order of the correlation measurement of the
$i$-th path. The normalized estimation precision with respect to
the resource $R$ is given by
\begin{equation}
\Delta t=\frac{1}{\sqrt{R}}\frac{\sqrt{\langle O^{2}\rangle-\langle O\rangle^{2}}}{|\frac{\partial\langle O\rangle}{\partial t}|}\sqrt{\sum_{i}Tr\{n_{i}\rho\}}.
\end{equation}
Here, suppose the illumination state to be a multimode squeezing state
$|\phi\rangle=\sum_{n}\beta_{n}|n,n...n\rangle$. For $N=3$ and $k_{1}=k_{2}=k_{3}=1$,
the estimation precision for the ``\emph{triple-pass}'' scheme
shows a maximum $\sqrt{3}$-fold enhancement over the ``\emph{single-pass}''
scheme when $t$ approaches unity.
\\

$^{\dagger}$These authors contribute equally.

\end{document}